\documentclass[fleqn,10pt]{wlscirep}
\usepackage[utf8]{inputenc}
\usepackage[T1]{fontenc}
\usepackage{bm}
\usepackage{amssymb,euscript}

\title{Nonlocal elastic metasurfaces: enabling broadband wave control via intentional nonlocality}

\author[1]{Hongfei Zhu}
\author[2,$\dagger$]{Timothy F. Walsh}
\author[2,$\dagger$]{Bradley H. Jared}
\author[1,*]{Fabio Semperlotti}
\affil[1]{Ray W. Herrick Laboratories, School of Mechanical Engineering, Purdue University, West Lafayette, Indiana 47907, USA}
\affil[2]{Sandia National Laboratories, Albuquerque, New Mexico 87185, USA}
\affil[*]{e-mail: fsemperl@purdue.edu}
\affil[$\dagger$]{Sandia National Laboratories is a multimission laboratory managed and operated by National Technology and Engineering Solutions of Sandia, LLC., a wholly owned subsidiary of Honeywell International, Inc., for the U.S. Department of Energy’s National Nuclear Security Administration. With main facilities in Albuquerque, N.M., and Livermore, C.A., Sandia has major R\&D responsibilities in national security, energy and environmental technologies, and economic competitiveness.}

\begin{abstract}
While elastic metasurfaces offer a remarkable and very effective approach to the subwalength control of stress waves, their use in practical applications is severely hindered by intrinsically narrow band performance. This work introduces the concept of intentional nonlocality as a fundamental mechanism to design passive elastic metasurfaces capable of an exceptionally broadband operating range.
The nonlocal behavior is achieved by exploiting nonlocal forces, conceptually akin to long-range interactions in nonlocal material microstructures, between subsets of resonant unit cells forming the metasurface. These long-range forces are obtained via carefully crafted flexible elements whose specific geometry and local dynamics are designed to create remarkably complex transfer functions between multiple units. The resulting nonlocal coupling forces enable achieving phase gradient profiles that are function of the wavenumber of the incident wave.The identification of relevant design parameters and the assessment of their impact on performance are explored via a combination of semi-analytical and numerical models. The nonlocal metasurface concept is tested, both numerically and experimentally, by embedding a total-internal-reflection design in a thin plate waveguide. Results confirm the feasibility of the intentionally nonlocal design concept and its ability to achieve a fully passive and broadband wave control.

\end{abstract}
\begin{document}
\flushbottom
\maketitle



\section{Introduction}
The concept of metasurface was originally pioneered in optics \cite{Yu,Ni,Grady,Pfeiffer,Aieta,Kang,Sun,Huang} and, shortly afterwards, extended to acoustics \cite{Li1,Li2,Mei,Tang,Yuan,Zhao1,Zhao2,yifan,Tang2,Li5,AcMs1,AcMs2}. Metasurfaces quickly gained popularity thanks to their ability to achieve wave front control via deep subwavelength artificial interfaces. This characteristic stands in stark contrast with traditional metamaterials whose ability to control incoming wave fronts is limited by the strict dependence between the size and number of unit cells, and the wavelength of the incoming wave. The most direct consequence of this physical constraint is the existence of a minimum dimension of a metamaterial slab (typically on the order of several wavelengths) that is necessary to control incoming waves. On the contrary, metasurfaces are very thin layers (typically a fraction of the dominant wavelength) that allow an abrupt control of the wave fronts, therefore offering a powerful alternative to circumvent a key practical limitation of metamaterials.

The response of a metasurface subject to an incoming wave is described by the Generalized Snell's Law (GSL) \cite{Yu,YuRev} which allows predicting directions of anomalous reflection and refraction across an interface characterized by a spatial phase gradient. It is indeed this capability to encode the most diverse phase gradient profiles within the metasurface that allows achieving unconventional effects including, but not limited to, wave front shaping \cite{Ni,Huang}, surface modes generation \cite{Sun}, and ultra-thin lenses \cite{Aieta,Kang}.

More recently, the concept of metasurface was also extended to elastodynamics \cite{ZhuPRL,EMS1,EMS2,EMS3,EMS4,EMS5,EMS6,EMS7} in an effort to achieve abrupt control of elastic waves in solid media. Initially, Zhu \textit{et al.} \cite{ZhuPRL} explored the behavior of an elastic metasurface in transmission mode and investigated the possibility of controlling the direction and the shape of a refracted wave front generated by an incoming planar wave. Later, the study was extended to show that metasurfaces could act as subwavelength barriers, therefore completely preventing the transmission of elastic waves across the interface \cite{TIR1,TIR2}. 
Despite the great success in the application of metasurfaces to wavefront control and shaping, their intrinsic narrowband performance has greatly limited their use in real-world applications. The limited operating frequency range is a direct consequence of the underlying design based on locally resonant unit cells. In passive metasurfaces, local resonances are used to achieve abrupt phase changes as propagating waves cross the interface. As the frequency of the incoming wave is detuned with respect to the operating frequency of the individual unit cells, the metasurface looses quickly its efficiency and ability to control the wave.

In order to address this important shortcoming of passive metasurfaces, researchers have explored the possibility to use active designs, involving piezoactuators and control logics, to extend the operating range. Chen \textit{et al}\cite{Chen_2018} introduced a programmable active elastic metasurface with sensing and actuating units, that achieved wave control functionalities over a broadband range of harmonic signals.
Despite the good success of the active approach, the overall increased complexity and the significant barriers in scaling the design to either larger structures or high intensity waves makes this approach not always viable. It follows that the ability to synthesize fully passive broadband designs still represents a major research endeavor and holds the potential to revolutionize metasurface-based devices. 

To address this major challenge, we introduce the concept of \textit{intentional non-locality}. 
The study and modeling of non-local system properties have been a long standing challenge in many areas of engineering and physics. While the general concept of non-locality holds a practically universal meaning, strictly connected to the idea of action at a distance, technical nuances can arise depending on the specific area of application. In classical mechanics of solids, the concept of non-locality implies that the local response of the medium at a target location (typically expressed in terms of a stress field) does not only depend on the state of the medium (typically expressed in terms of a strain field) at that same location but also from the state of other distant locations. The ensemble of these distant points, whose state affects the response at the target location, forms a surface (in 2D) or a volume (in 3D) known as the area, or volume, of influence. In elastodynamics, the concept of nonlocality results in a dependence of the stress (or, equivalently, the strain) field on the wavenumber\cite{NOWINSKI1,Nowinski2,ERINGEN1,ERINGEN2}. Note that the typical nonlocal character of many natural materials is very subtle and, in most applications, it is considered as a higher order effect.

The concept of nonlocality has also found applications in electromagnetism. Compared to the mechanical properties, nonlocality has a much more pronounced effect on electromagnetic properties \cite{NLO1,NLO2,NL03}. This is also the reason why nonlocal designs have found natural applications on electromagnetic metasurfaces \cite{NCEMMS2,NCEMMS1}, metagratings \cite{NCEMMS3,NCEMMS4,NCEMMS5} and metamaterials \cite{NLEMM1,NLEMM2,NLEMM3}. More recently, this concept was also extended to acoustic metasurfaces \cite{NCAM}.

To-date, no attempts were made to develop nonlocal elastic metasurfaces. In part, this technology gap follows from the weak nonlocal mechanical effect in most natural materials, as previously mentioned. 
In this study, we propose the concept of nonlocality that is intentionally introduced into the design of an elastic metasurface in order to extend its operating range by leveraging wavenumber-dependent mechanical properties. More specifically, starting from a classical locally resonant elastic metasurface design\cite{ZhuPRL}, we exploit the concept of \textit{action at a distance} by coupling selected unit cells via carefully designed connecting elements. These elements are conceived explicitly to create wavenumber-dependent coupling forces that mimic, at the macroscale, the nonlocal effect typical of nonlocal forces in elastic continuum microstructures.

In this study, the concept of intentional nonlocality will be illustrated by applying it to a total-internal-reflection type of metasurface (TIR-MS) \cite{TIR1,TIR2}, which was chosen as basic benchmark system. Note that, despite this choice, the proposed concept is extremely general and applicable to any type of metasurface. A specific semi-analytical transfer-matrix model was developed to study the effect of the nonlocality on the dynamic behavior of the metasurface and to facilitate the synthesis of practical physical designs for numerical and experimental testing. Then, the nonlocal TIR-MS concept was embedded in a thin plate waveguide and its performance assessed via a combination of numerical and experimental tests.

\section{Fundamentals of local elastic metasurface design}

In order to facilitate the understanding of the nonlocal design, this section provides a brief overview of the basic design principles of a classical passive elastic metasurface\cite{ZhuPRL}, which in the following will be referred to as the \textit{local} design. In the classical problem of wave transmission across an interface between dissimilar materials, the angle of refraction is entirely determined by the impedance mismatch between the two materials and it is carefully described by the Snell's law. For a given wavelength, the angle of refraction is fixed once the two materials are chosen. However, when in presence of a metasurface that encodes a prescribed spatial phase gradient (along the metasurface direction), the refracted angle is no longer controlled only by the impedance mismatch between materials but also by the phase gradient itself. In fact, metasurfaces are typically employed to control the reflection and refraction of waves between halves spaces made of the same material (hence without an intrinsic impedance mismatch).

\begin{figure}[h]
	\centering
	\includegraphics[scale=0.85]{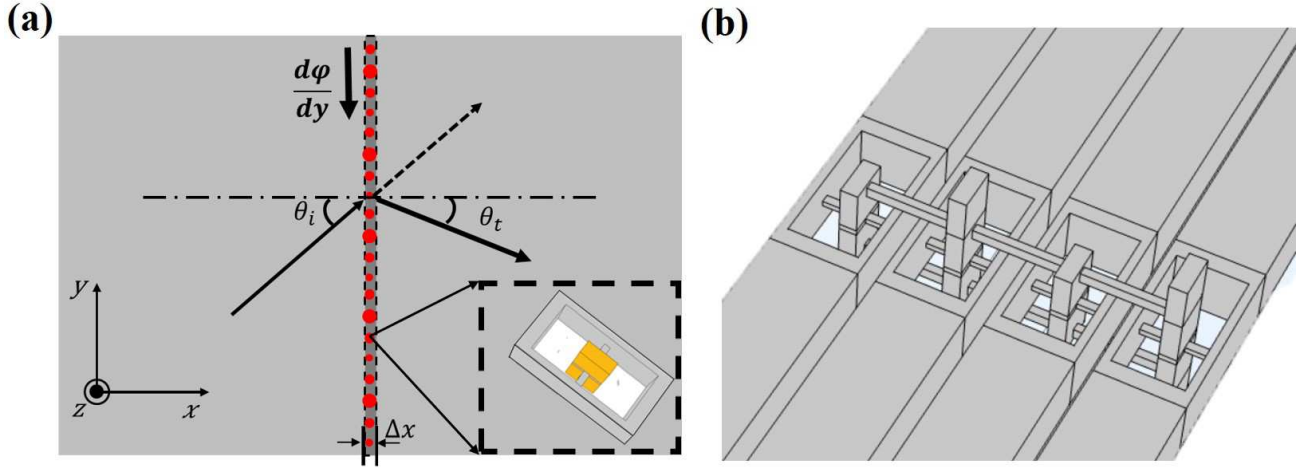}
	\caption{(a) Schematic view of an elastic waveguide with an embedded local metasurface. Some of the fundamental parameters such as the angle of incidence $\theta_i$, the angle of anomalous refraction $\theta_t$, and the phase gradient $\frac{\rm d\phi}{\rm d \it y}$ are shown. The inset also shows a schematic view of a typical locally resonant unit from which the metasurface is assembled. (b) Schematic illustration introducing the concept of intentional nonlocality realized via flexible elements connecting different unit cells. } \label{figa1}
\end{figure}

When in presence of an interface with a spatial gradient profile (Fig.~\ref{figa1}a), the direction of the refracted wave front is obtained from the Generalized Snell's Law (GSL)\cite{Yu}:
\begin{equation}
\frac{\rm sin(\theta_{\it t})}{\lambda_t}- \frac{\rm sin(\theta_{\it i})}{\lambda_i}=\frac{1}{2\pi}\frac{\rm d\phi}{\rm d \it y}\label{eqn1}
\end{equation}
where $\rm d\phi /d \it y$ is the phase gradient, while $\theta_i$ and $\theta_t$ indicate the directions of the incident and refracted waves, respectively.

In passive elastic metasurfaces, the phase gradient is often realized by using a periodic array of locally resonant cells providing a linearly-varying, step-like phase modulation covering the range [0,2$\pi$]. 
Figure~\ref{figa1}a shows a conceptual view of a typical local elastic metasurface as well as a zoomed-in view of the locally resonant unit cell. The latter consists of a hollow frame with an internal resonator realized via a small mass attached to two slender beams connected to the frame. Previous studies\cite{TIR1,TIR2} showed that the dynamics of the resonator can be easily controlled in order to match specific phase profiles and design constraints while limiting the fabrication complexity.

While, in the local design, the desired phase gradient profiles can be efficiently achieved at a single target frequency, it is not possible to maintain the target phase gradient over a wide frequency range. This limitation is a direct consequence of the fundamental operating mechanism of the metasurface based on local resonances. A possible approach to overcome this limitation could include the design of a multi-resonant unit cells (e.g. using internal multi-degree-of-freedom resonators) within the basic local metasurface design. The increased number of design parameters would improve the ability to tune the phase gradient profile versus frequency. Although conceptually possible, this approach would likely result in a very complex design of the unit cells and in an operating range still limited by the number of local resonances available per unit cell.

\section{Basic concept and mechanics of the nonlocal supercell}

The concept of nonlocal metasurface presented in this study is based on the fundamental idea of a nonlocal supercell. A nonlocal supercell consists of a prescribed number of locally resonant units whose internal resonators are coupled to each other via specifically designed flexible links. The role of these links is to provide, at macroscopic level, an action at a distance that is conceptually equivalent to the role that nonlocal forces play in nonlocal microstructures. The main objective of these long-range connections is to enrich the dynamics of each individual unit cell which is now influenced also by the dynamics of distant cells. The most immediate effect of these coupling forces is the generation of a wavenumber-dependent dynamics, for each individual cell, that can be made as complex as needed to achieve a target phase profile over a broad range of operating frequencies.
Once the nonlocal supercell design is available, the nonlocal metasurface can be assembled by realizing a periodic distribution of identical nonlocal supercells.

For elastic metasufaces, a possible implementation of the nonlocal design is shown in Fig.~\ref{figa1}b. The design includes flexible beam-like links that connect the local resonators of multiple individual units. As explained in detail in the following, in order to obtain complex phase profiles over an extended frequency range, the connecting elements must exhibit an elaborate local dynamics that can be obtained in a variety of ways, including proper choices of the cross-sectional properties, materials, and local resonances.
Within this general design framework, the synthesis of the links' final configuration requires the formulation and solution of an optimization problem. The objective of the optimization procedure is to select the links' physical parameters capable of producing the desired nonlocal forces that ultimately enable maintaining the target phase profile. 
	
\begin{figure}[h]
	\includegraphics[scale=0.85]{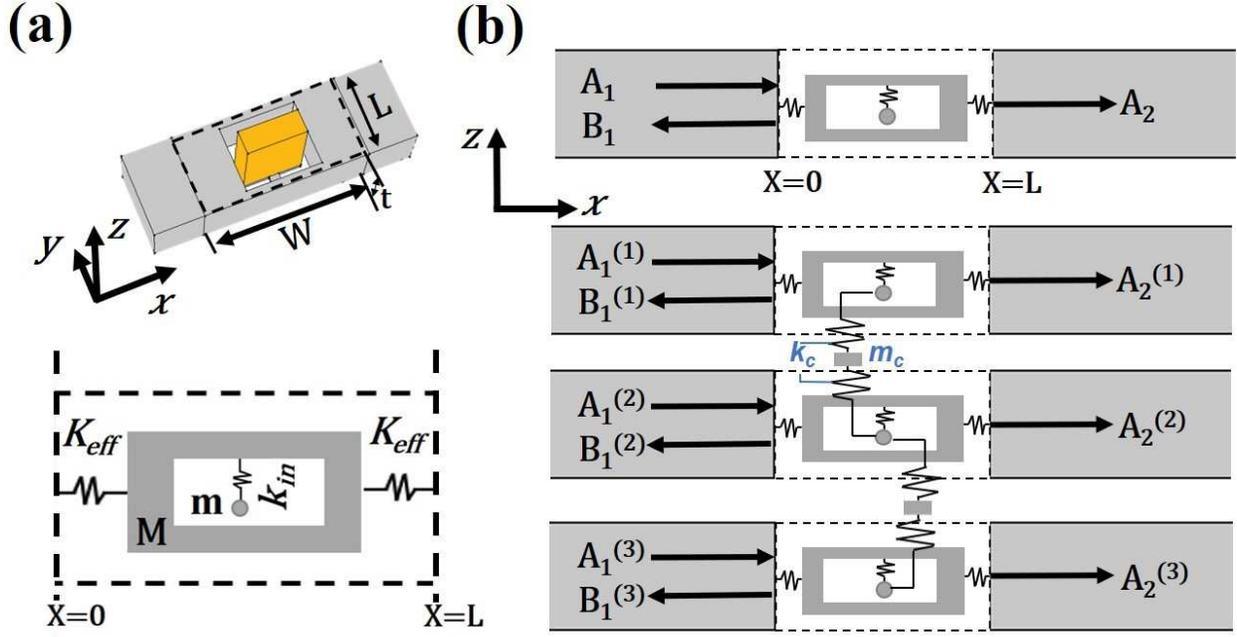}
	\caption{Schematics illustrating the conversion of the unit cell model from a continuum to a discrete representation. This conversion is at the basis of the development of the transfer matrix model. (a) (top) Schematic of a typical locally resonant unit cell, and (bottom) its lumped parameter counterpart represented by a mass-in-mass-spring model. (b) (top) Schematic showing the integration of the lumped unit cell model within the elastic waveguide as well as the amplitude of the incident ($A_1$), reflected ($B_1$), and transmitted ($A_2$) waves. (bottom) Schematic showing the integration of local resonant units to form a nonlocal supercell. The connecting links are represented by lumped elements having stiffness $k_c$ and mass $m_c$. The superscript $(\square)$ in the amplitude coefficients represents the units number.} \label{fig1}
\end{figure}

Dedicated analysis tools are needed in order to simulate the response of these nonlocal systems, specifically, to understand the effect of the design parameters on the physical behavior of the supercell and to allow the synthesis of physically realizable designs.
The overall modeling approach follows the following strategy. The nonlocal supercells are initially modeled using simplified lumped parameter (e.g. mass-spring) systems that allows formulating the system dynamics via transfer matrix (TM) method. In the TM approach, the resonant unit is represented via a lumped parameter system, conceptually shown in Fig.~\ref{fig1}a, which allows capturing the relevant dynamics while requiring a reduced set of system parameters. This simplified approach is possible because the metasurface operates within a deep-subwavelength regime (i.e. the dominant wavelength is much larger than the characteristic unit cell dimension). In addition, the long-range coupling forces between local unit cells can be simulated by means of mass-spring connections having frequency-dependent properties. Once the unit cells are coupled to form the \textit{nonlocal} supercell, the links' parameters can be tuned to achieve a frequency-dependent phase profile. 

\subsection{Mathematical model of the nonlocal metasurface}
In this section, we present the main highlights of the mathematical model developed to design and simulate the dynamics of the nonlocal metasurface. Complete details can be found in the Supplementary Information (SI \cite{SI}). As previously mentioned, the dynamic behavior of a unit cell is estimated by means of a discrete (mass-in-mass)-spring model via a transfer matrix (TM) approach. Figure~\ref{fig1}a shows both the continuum and the discrete representations of the unit cell and it clarifies the role of the lumped parameters. In the discrete model, the unit cell is made of a mass $M$ that captures the cell's frame, a mass $m$ that represents the internal resonator, and a spring $k_{in}$ that connects the resonator to the frame. The resulting model is a classical mass-in-mass system\cite{lUMPEDMODEL1,lUMPEDMODEL2}. The unit cell is connected to the exterior boundaries (representing the waveguide) by effective springs having stiffness $K_{eff}$.

Assuming time-harmonic response at an angular frequency $\omega$, the governing equations of the (mass-in-mass)-spring units are given by,
\begin{subequations}
	\begin{gather}
	M\frac{\partial^2w^M}{\partial t^2}=K_{eff}(w_0+w_L)-2K_{eff}w^M+k_{in}(w^m-w^M)\\
	m\frac{\partial^2w^m}{\partial t^2}=k_{in}(w^M-w^m)\\
	F_0=K_{eff}(w^M-w_0)\\
	F_L=K_{eff}(w_L-w^M)
	\end{gather}
\end{subequations}
where $w$ and $F$ represent the boundary displacement and shear force, while the subscript $0$ and $L$ indicate that the variable is taken at either the input or the output boundary, respectively and the superscript $M$ and $m$ of $w$ indicate the displacement of the frame mass $M$ and  the internal resonator $m$. $L$ is the length of the unit cell. Solving the above system of equations, the quantities at the output boundary can be expressed as a function of the same quantities at the input boundary, 
\begin{equation}
\begin{bmatrix}
w_L\\F_L\end{bmatrix}=T\begin{bmatrix}
w_0\\F_0\end{bmatrix}
\end{equation}
where the transfer matrix $T$ is given by,
\begin{equation}
T=\begin{bmatrix} 
T_{11} & T_{12} \\[3ex]
T_{21} & T_{22} 
\end{bmatrix}=\begin{bmatrix} 
\frac{K_{eff}-\omega^2(M+\frac{k_{in}m}{k_{in}-m\omega^2})}{K_{eff}} & \frac{2K_{eff}-\omega^2(M+\frac{k_{in}m}{k_{in}-m\omega^2})}{K^2_{eff}} \\
-\omega^2(M+\frac{k_{in}m}{k_{in}-m\omega^2})  & \frac{K_{eff}-\omega^2(M+\frac{k_{in}m}{k_{in}-m\omega^2})}{K_{eff}}  
\end{bmatrix}\label{eq1}
\end{equation}

The TM model can then be extended to simulate the nonlocal supercell in which multiple units are coupled via connecting elements, as shown in Fig.~\ref{fig1}b. While the physical design of the connecting beams will require additional steps, at this stage their dynamics is captured via discrete spring-mass elements that model the effective stiffness $k_c$ and mass $m_c$ of the link. As an examples, considering the case where two units are coupled together by a link, the equations of motion for the unit I or II are given by:
\begin{subequations}
	\begin{gather}
	M^{(\square)}\frac{\partial^2w^{M^{(\square)}}}{\partial t^2}=K^{(\square)}_{eff}(w^{(\square)}_0+w^{(\square)}_L)-2K^{(\square)}_{eff}w^{M^{(\square)}}+k^{(\square)}_{in}(w^{m^{(\square)}}-w^{M^{(\square)}})\\
	m^{(\square)}\frac{\partial^2w^{m^{(\square)}}}{\partial t^2}=k^{(\square)}_{in}(w^{M^{(\square)}}-w^{m^{(\square)}})+k_c(w_{mc}-w^{m^{(\square)}})\\
	F^{(\square)}_0=K^{(\square)}_{eff}(w^{m^{(\square)}}-w^{(\square)}_0)\\
	F^{(\square)}_L=K^{(\square)}_{eff}(w^{(\square)}_L-w^{m^{(\square)}})
	\end{gather}
\end{subequations}
where the superscript $(\square)$ can be either I or II and it refers to the different unit cells in the nonlocal supercell, and the basic units are coupled by the connecting beam whose governing equation is given by,
\begin{equation}
m_c\frac{\partial^2w_{mc}}{\partial t^2}=k_c(w^{m^{(I)}}+w^{m^{(II)}})-2k_cw_{mc}
\end{equation}
$m_c$ and $k_c$ are the effective parameters characterizing the link in the form of a spring with mass (as shown in Fig.~\ref{fig1}b), and $w_{mc}$ is the displacement of the lumped link.

By solving the above set of differential equations, the response at the outer boundary of the coupled supercell can be expressed as a function of the quantities at the input boundary using a nonlocal transfer matrix $\tilde{T}$ ($4\times4$ in size),  
\begin{align}
& 
\begin{bmatrix}w^{(I)}_L\\f^{(I)}_L\\w^{(II)}_L\\f^{(II)}_L\end{bmatrix}=\tilde{T}\begin{bmatrix}
w^{(I)}_0\\f^{(I)}_0\\w^{(II)}_0\\f^{(II)}_0\end{bmatrix}
\end{align}
Once the nonlocal transfer matrix $\tilde{T}$ is determined, the boundary response of each unit can be further related to the incident, reflected, and refracted waves in order to extract both the phase and the transmission coefficients of the non-local supercell. More details on the methodology as well as a practical example of the simulation of a nonlocal supercell can be found in SI\cite{SI}. Results also include a direct comparison with the numerical predictions obtained from a three-dimensional finite element model. Despite the several simplifications introduced in the TM approach, the two models are in good agreement hence indicating that the simplified approach is perfectly capable of capturing the behavior of the unit cells, the connecting links, and the metasurface as a whole.

\section{Nonlocal total-internal-reflection elastic metasurface (NL-TIR-MS)}

In this section, we apply the proposed methodology to design a nonlocal total-internal-reflection metasurface~\cite{TIRAPL} (NL-TIR-MS) embedded in a plate waveguide. We recall that a local TIR-MS \cite{TIRAPL, TIR2} consists in a metasurface that is specifically designed to exhibit a phase gradient profile exceeding the requirements for total internal reflection. Under these conditions, incoming waves with any arbitrary angle of incidence (but in the neighborhood of the target operating frequency, i.e. the resonance frequency of the local unit cells) will not be able to cross the interface and will be reflected back. 
Note that the selection of the TIR-MS as benchmark system to test the nonlocal concept does not limit the generality of the results. Similar performance can be envisioned also for transparent metasurfaces.

Both the metasurface and the host waveguide have thickness $t=6.35$ mm. The target operating bandwidth is set to $\Delta f=[4.2,5.7]$ kHz. Note that this selection of the operating range is somewhat arbitrary and it was chosen exclusively for convenience of the experimental measurement. The nonlocal approach \textit{per se} is not limited to a specific frequency range. 
The overall dimensions of the basic unit are $L(=20 \ mm)\times W(=10 \ mm )\times t(=6.35\ mm)$, as shown in Fig.~\ref{fig1}a. Recall that, in order to ensure TIR conditions\cite{TIR1}, the phase shift gradient should always exceed the critical value $\rm d\phi /  \rm d \it y$ $= 4\pi/\lambda$, and the transmission coefficients of each unit should have a comparable value. Assuming that the supercell contains $n$ basic units to cover the $2\pi$ range, then it must be $\frac{2\pi}{n W} \geq 4\pi / \lambda$ which results in a condition on the number of units $ n \leq \frac{\lambda}{2 W} $. Over the selected frequency range, the wavelength $\lambda$ approximately varies between $12.3 W$ and $10.1 W$, hence a nonlocal supercell of $n=3$ units would fulfill the TIR criterion.

Assuming three basic units, the synthesis of the physical design (that is the link that exhibits the desired effective dynamic properties predicted by the TM method) requires an optimization approach. To achieve a physical design, the procedure is separated into three main steps: 1) select the fundamental unit cells, 2) identify the necessary nonlocal interaction force versus frequency profile, and 3) synthesize the physical design of the links capable of realizing the target nonlocal force profile. These three steps are further explained in the following.

\begin{figure}[h]
	\centering
	\includegraphics[scale=0.85]{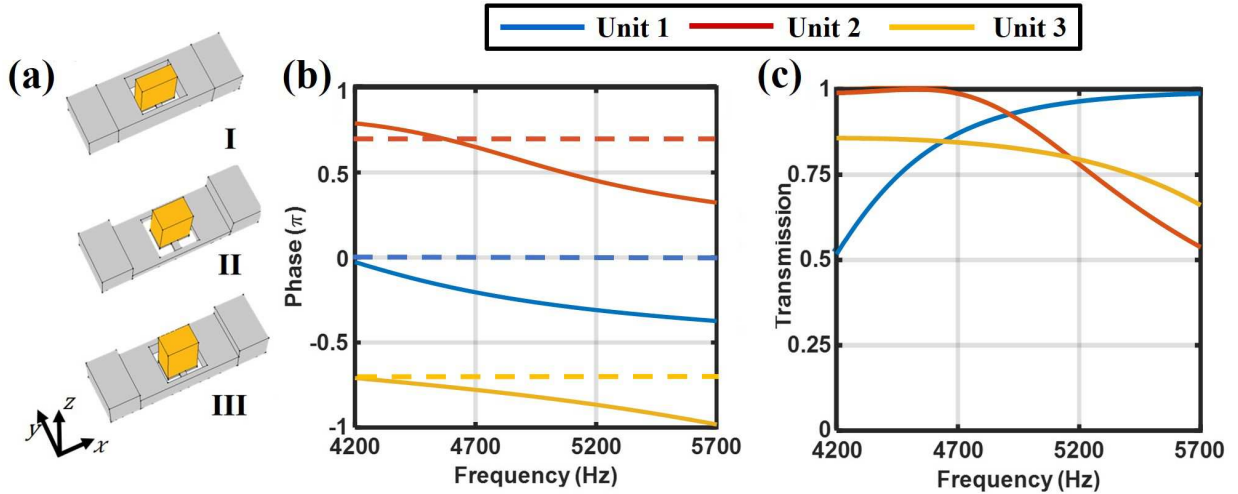}
	\caption{ (a) Schematics of the three selected locally resonant unit cells that will form the final nonlocal TIR-MS. (b) and (c) shows both the transmission phase and the transmission amplitude versus frequency for the three units. } \label{fig4}
\end{figure}

\textbf{Step 1:} the first step consists in selecting the physical design of the locally resonant unit cells. Note that, contrarily to the local design \cite{ZhuPRL} which requires setting the fundamental resonance of the unit cells at the target frequency, in the nonlocal design the individual unit cells do not need to be designed for a specific frequency. In fact, it is even preferable not to start with units operating at the same frequency because this would bias the dynamic behavior of the nonlocal metasurface in favor of a single frequency. The most direct effect of this bias would be the need for extreme values of the effective properties (e.g. either zero or negative mass and stiffness values) required from the links in order to guarantee broadband performance. On the other side, an appropriate selection of the basic units can facilitate achieving a broadband design. In general, it was found to be a good design rule for the nonlocal metasurface to select units whose phase contribution (compared with the adjacent units) is close to $2\pi/n$ (where n=3 in the present case) over the selected frequency range. This first step of the design can be performed by acting on the basic geometric parameters of the unit cells in order to tune their local resonances and, consequently, the phase profile. For the example presented in this study, the three selected units are shown in Fig.~\ref{fig4}a. Details on the geometric design can be found in SI\cite{SI}. Their amplitude and phase responses were extracted using the discrete TM model and the results are summarized in Fig.~\ref{fig4}b and c. It can be seen that, although the phase profiles of the three units are slowly varying over the selected frequency range $0$, $2\pi/3$ and $-2\pi/3$, the corresponding phase jumps between adjacent units (Fig.~\ref{fig4}b) are always in the neighborhood of the target value.

\begin{figure}[h]
    \centering
	\includegraphics[scale=0.75]{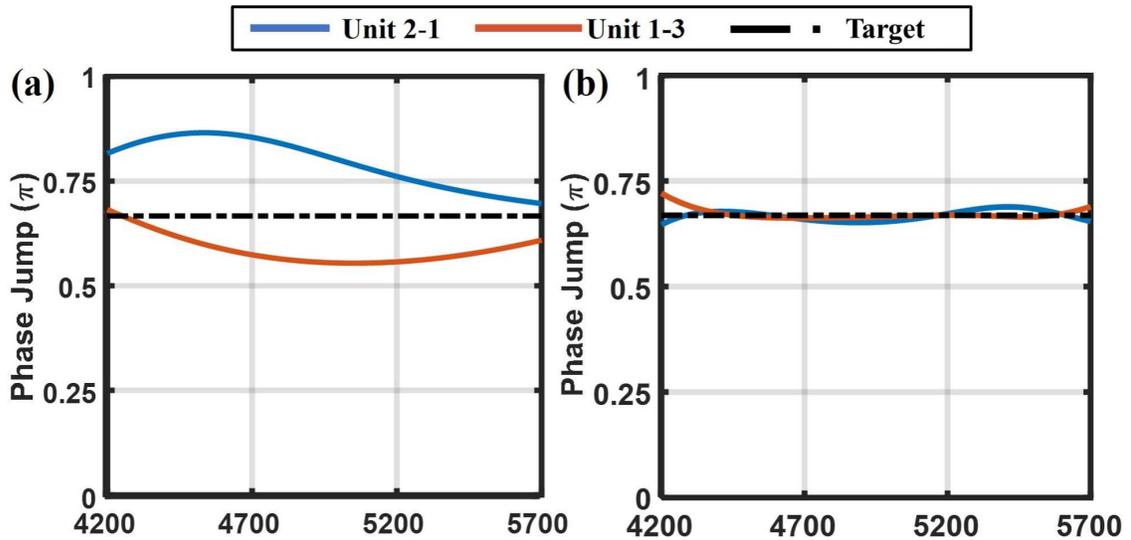}
	\caption{(a) The phase gradient of the local supercell without coupling elements. (b) The predicted phase gradient from the lumped nonlocal supercell model after the application of the optimized coupling elements.} \label{fig5}
\end{figure}

\begin{figure}[h]
    \centering
	\includegraphics[scale=0.85]{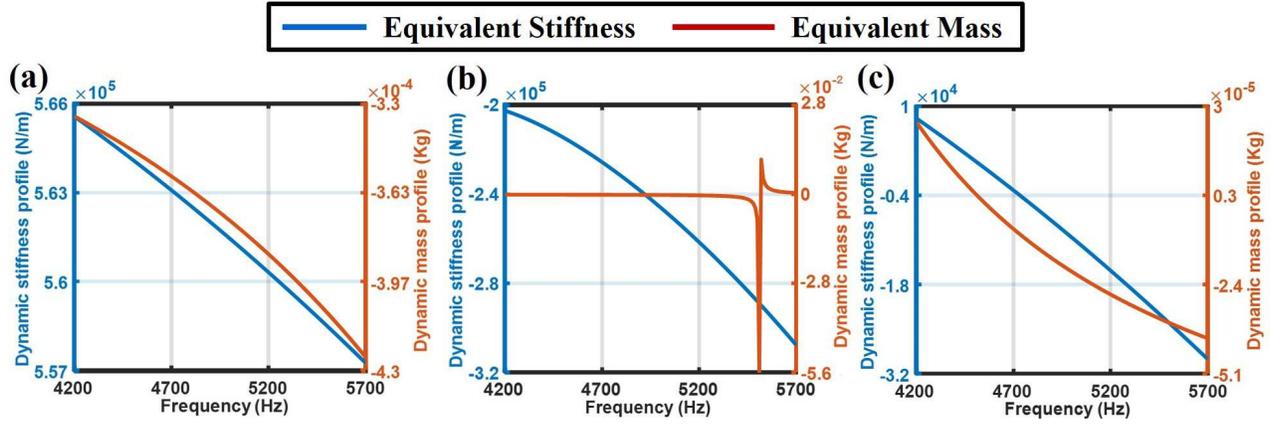}
	\caption{The required dynamic stiffness and dynamic mass profiles necessary to achieve the broadband TIR effect. Results are presented for the three coupling beams: (a) beam I (connecting unit 1 to 2), (b) beam II (connecting unit 2 to 3), and (c) beam III (connecting unit 1 to 3).} \label{fig6}
\end{figure}

\textbf{Step 2:} once the fundamental resonant units forming the supercell are selected, the next step requires the identification of the nonlocal (coupling) interaction forces needed to achieve the target performance of the TIR-MS over the selected operating range. Equivalently, the former objective can be stated by saying that the frequency-dependent effective properties (i.e. $m_c(\omega)$ and $k_c(\omega)$) that identify the dynamics of each connecting element must be determined such that the phase jump between adjacent units is maintained at the constant value $2\pi/3$. Note that, although the amplitude of the response of each unit should also be maintained to comparable levels to each other (a condition that guarantees the transmission coefficients of each unit to be equivalent), for the TIR-MS design the ability to guarantee a constant phase gradient poses a more stringent requirement. In other terms, the ability to yield the desired phase gradient is more important than matching the response amplitude at the unit level.
The phase response of the local and nonlocal three-unit-supercell are presented in Fig.~\ref{fig5}a and b, respectively. Results show that the nonlocal design is capable of producing a reasonably constant phase at the desired value $2\pi/3$ (see also SI\cite{SI}). The corresponding effective properties characterizing the three coupling elements are shown in Fig.~\ref{fig6}a, b, and c, respectively. A visual analysis of these wavenumber-dependent functions allows making an important consideration. The dynamic mass profile of unit cell \#2 (Fig.~\ref{fig6}b) shows a sharp variation around 5.4 kHz. While, at this stage, we are dealing with synthetic coupling forces (i.e. effective forces not yet associated with a physical design of the link) this result suggests that to be able to achieve such a sharp change in the dynamic properties of the link a local resonance should be exploited. It is also worth noting that, being this the result of an inverse problem, the set of force profiles discussed above is only one of the many possible solutions. 

\begin{figure}[h]
    \centering
	\includegraphics[scale=0.85]{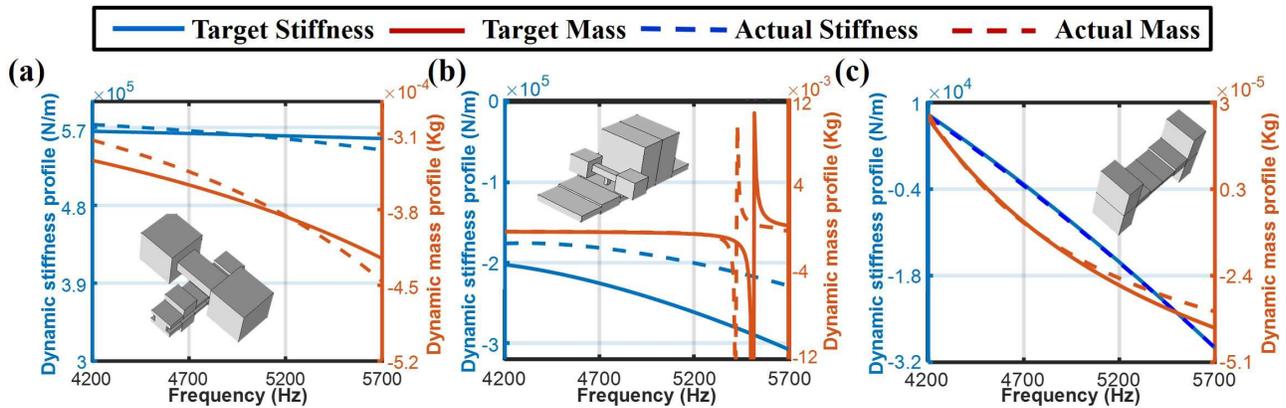}
	\caption{Comparison of the dynamic stiffness and dynamic mass profiles provided by the synthetic (dashed lines) and the physical (solid lines) designs. The insets in each figure show the corresponding physical design of the beam connectors and local resonators: (a) beam I, (b) beam II, and (c) beam III. } \label{fig7}
\end{figure}

\textbf{Step 3:} once the necessary dynamic mass and stiffness profiles of the connecting elements are identified, the third and last step consists in converting these synthetic long-range transfer functions into physically realizable designs of the three coupling elements. To simplify the design process and avoid the need for topology optimization, we assume that the basic geometry of the connecting elements is fixed to be a thin rectangular beam having piece-wise constant thickness. While changes in thickness can help controlling the flexibility and the global dynamics of the link, it was already pointed out that sharp variations in the coupling forces requires local resonances. Hence, the beam connectors will also include attached resonators tuned at selected frequencies. Also in this case, in order to avoid the use of topology optimization, we assumed these resonator to be in the form of a cantilever dumbbell. 
Based on the above assumptions, the topology of the links is entirely defined, so that the profiles of the effective properties (or, equivalently, of the nonlocal coupling forces) can be matched by performing a simple parameter optimization on the geometric design variables. This step was performed using a commercial finite-element-based optimizer (COMSOL Multiphysics). The resulting physical configurations of the three coupling beams are shown in Fig.~\ref{fig7} and their dynamic responses are evaluated in Fig.~\ref{fig7}a, b and c (dashed lines), and compared with the target profiles (solid lines). Overall, the dynamic properties of the three beams are in good agreement with the target profiles. Note that this study focuses on presenting and validating the concept of intentionally nonlocal metasurfaces, not on the evaluation of optimal designs. Hence, it is merely highlighted here that a concurrent optimization approach could provide better match of the nonlocal coupling terms. 

\section{Numerical Results}

\begin{figure}
	\includegraphics[scale=0.85]{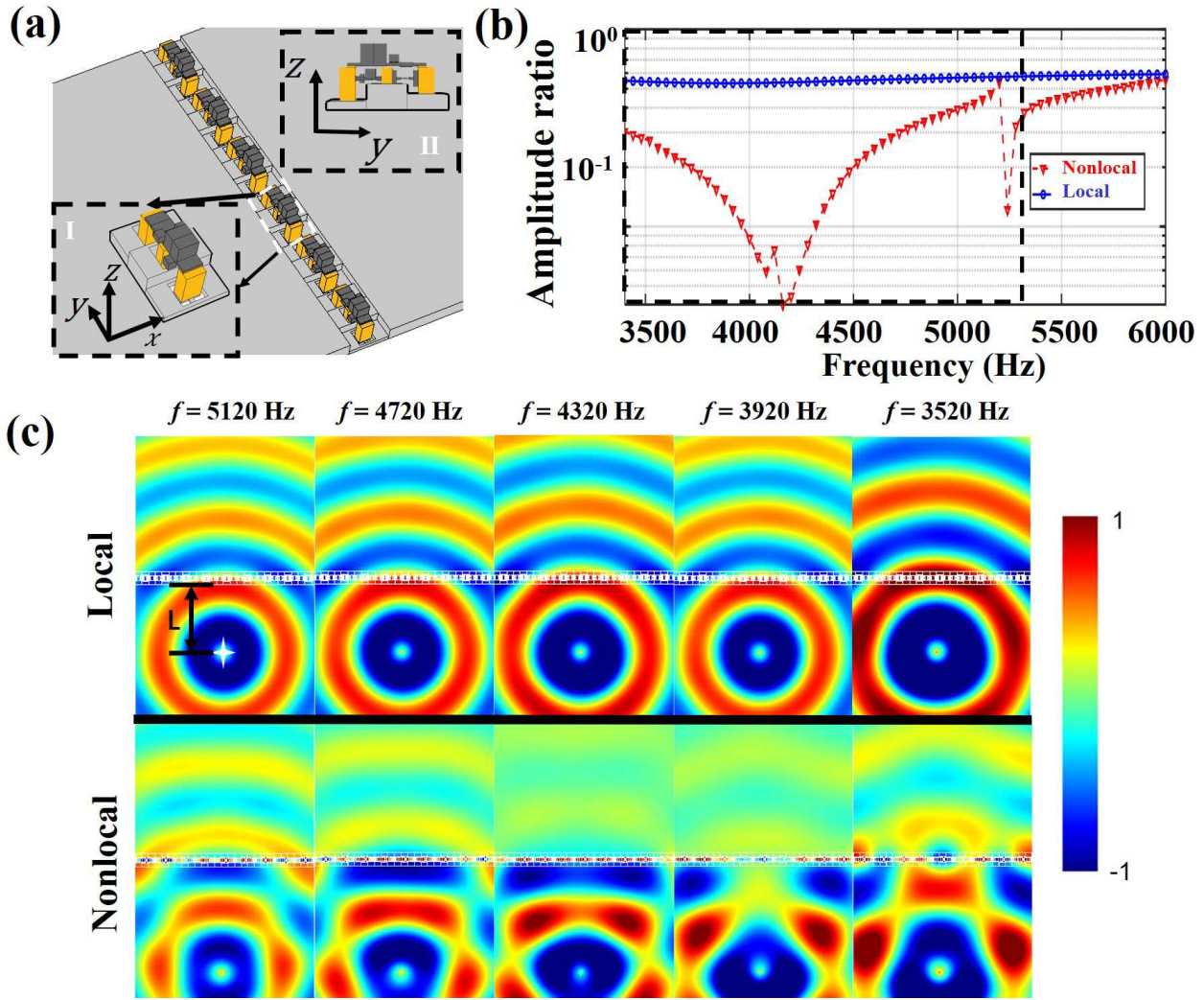}
	\caption{Numerical simulations obtained from a full 3D model and showing the performance of the nonlocal TIR metasurface. (a) Schematics showing the 3D model of the assembled nonlocal TIR-MS embedded in a thin waveguide. The insets show the isometric (I) and the side (II) view of the nonlocal connectors mounted on the resonating cores. Different components are marked by colors: basic unit cells (yellow) and connecting links (dark grey). (b) Spatially-averaged amplitude ratio of the out-of-plane component of the velocity  obtained from the response of the waveguide before and after the MS interface. Both local and nonlocal designs are compared. The wide dip observed in the nonlocal MS response (highlighted by the dashed black box) indicates the broadband wave blocking effect. (c) full field simulations showing the out-of-plane particle displacement at five selected frequencies within the nonlocal MS operating range. The white star marker shows the location of the point load source excitation.} \label{Fig8}
\end{figure}

The previous section presented the physical design of the connecting elements capable of guaranteeing to the metasurface a broadband TIR effect. Based on this physical design, a complete 3D model of the nonlocal metasurface (Fig.~\ref{Fig8}a) embedded in a flat plate was assembled. Note that the nonlocal metasurface was simply obtained by periodically repeating the basic nonlocal supercell (see inset I in Fig.~\ref{Fig8}a) along the interface direction. The side view of the non-local supercell ($y-z$ plane) is also provided (see inset II in Fig.~\ref{Fig8}a) in order to clarify the connection strategy between units. Further details on the connection scheme can be found in \textbf{SI}\cite{SI}.

The waveguide with the embedded nonlocal metasurface was modeled via finite element analysis using a commerical software (Comsol Multiphysics). The metasurface was excited by a point load source located at a distance of 0.12m from the plane of the metasurface (see the white star marker in Fig.~\ref{Fig8}c), and acting in the out-of-plane $z-$direction, hence generating predominantly $A_0$ guided modes. Perfectly matched layers (PML) were applied at all the boundaries to eliminate the effect of unwanted reflections. In order to assess the effect of the nonlocal metasurface on the propagating wave, the model of the equivalent local metasurface design was developed and used as a reference. This model consisted in the same set of resonant unit cells as those used in the nonlocal TIR-MS design but without the nonlocal connections and the large masses (drawings provided in SI\cite{SI}). 

Figure~\ref{Fig8}b shows the comparison of the spatially-averaged amplitude ratios calculated from the transverse particle displacement ($z-$direction) before and after the MS. Both the reference (local) MS and the NL-TIR-MS are shown. The large reduction in the amplitude ratio of the NL-TIR-MS metasurface (marked by the dashed black box in Fig.~\ref{Fig8}b) compared to the reference waveguide is a clear indicator of the broad operating range of the proposed nonlocal design. Full field numerical results are shown in Fig.~\ref{Fig8}c for five selected frequencies inside the operating bandwith. The results are still presented in terms of out-of-plane particle displacement of the $A_0$ wave field. The direct comparison of these wave patterns shows that, although a small fraction of the incident wavefront can still be transmitted through the metasurface, the nonlocal design is capable of producing a drastic reduction of the transmitted wave intensity over a broad frequency range.
In order to put these results in perspective, note that the numerically calculated bandwidth for the nonlocal design (approximately $\Delta f=[3.5,5.1]$ kHz) corresponds to a $37.2\%$ relative bandwidth with respect to its center frequency $f_0=4.3$ kHz. Compared to previous studies on local TIR-MS\cite{TIR1,TIR2} where available data showed approximately a 250 Hz bandwidth (only $4.54\%$ of the center frequency $f_0$=5.5 kHz), the nonlocal design provides a range that is approximately eight times larger.

\section{Experimental Validation}

In order to validate the concept of intentional nonlocality as a viable approach to the synthesis of passive broadband metasurfaces and to confirm the performance of the NL-TIR-MS, we built an experimental testbed following the configuration that was numerically studied in Fig.~\ref{Fig8}a. The plate waveguide and the supporting frames of each resonant unit cell were made out of aluminum and fabricated via computer controlled machining (CNC). The nonlocal connecting links (including the three resonating masses) were made out of 316L stainless steel and fabricated via additive manufacturing techniques (i.e. metal printing). The links were later glued on the corresponding unit cells slots realized in the waveguide. Further details on the fabrication process are discussed in SI\cite{SI}. 

\begin{figure}[ht]
    \centering
	\includegraphics[scale=0.7]{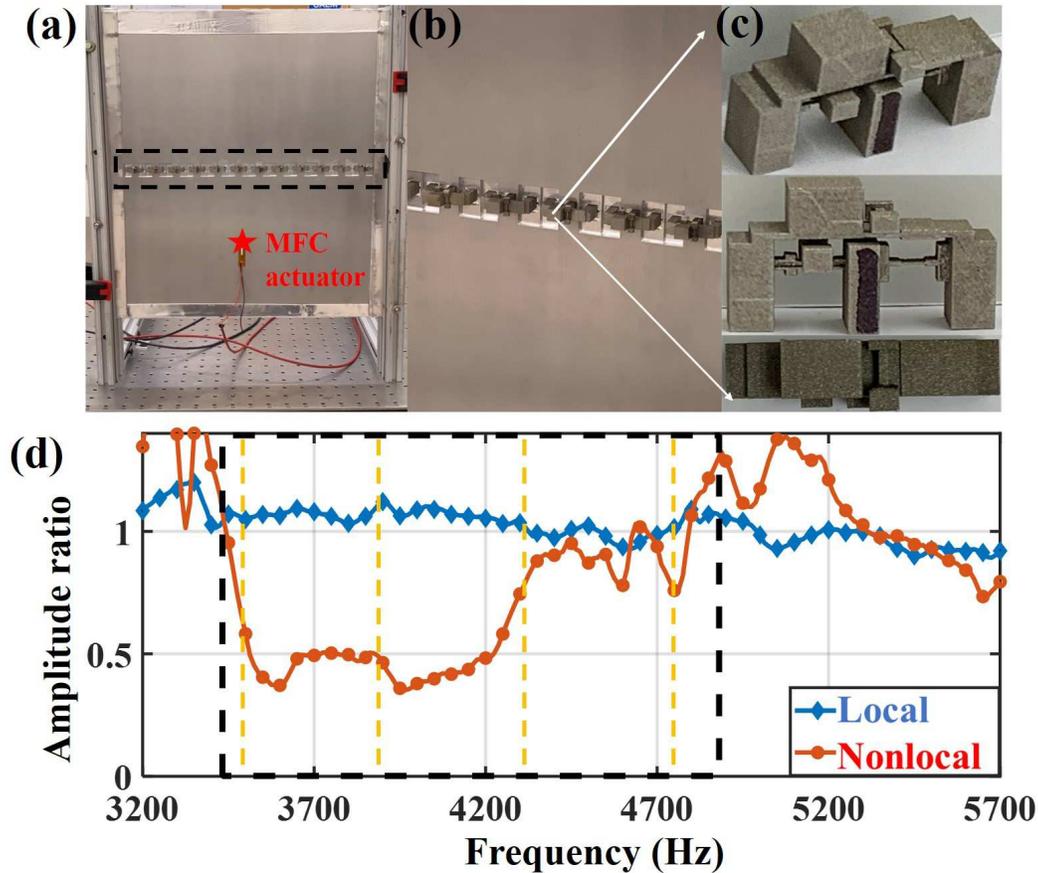}
	\caption{Experimental setup consisting of a thin plate waveguide with an embedded nonlocal TIR-MS. The waveguide was inserted into an aluminum frame providing clamped boundary condition on two sides of the waveguide. The excitation was provided by a MFC patch glued on the surface (see red star marker). (a) The front view of the experimental test bed. (b) Zoomed-in view of the nonlocal TIR-MS. (c) 3D metal printed nonlocal beams mounted on the the resonating masses. Three views are shown (Iso, Front, Top) moving from the top to the bottom panel. The connectors are later glued in place on the corresponding unit cells realized in the supporting waveguide. (d) The spatially-averaged amplitude ratio obtained from the measured out-of-plane velocity field before and after the MS. The comparison between the local and the nonlocal design clearly shows the broadband nature of the latter highlighted by the large drop in the amplitude ratio (see dashed black box).} \label{fig10}
\end{figure}

The test setup is shown in Fig.~\ref{fig10}. Figure \ref{fig10}a provides a front view of the setup consisting in a plate waveguide with the embedded nonlocal MS, while Fig.~\ref{fig10}b shows a closeup view of the metasurface. In Fig.~\ref{fig10}c, from top to bottom, the three panels show different views (Iso-, Front- and Top-) of the nonlocal links assembled on the three resonant masses. The plate waveguide was mounted in an aluminum frame while a single Micro-Fiber-Composite (MFC) actuator was glued on the surface of the plate in order to apply the external excitation (see the red star marker in Fig.~\ref{fig10}a). Viscoelastic tapes were applied all around the boundary of the plate so to reduce back-scattering. The response of the plate was measured under white noise excitation (bandwidth 3-6 kHz) by using a scanning laser Doppler vibrometer. Both the local (reference) and the nonlocal configurations were tested.

Figure~\ref{fig10}d presents the comparison of the spatially-averaged amplitude ratios collected from the measured out-of-plane particle velocity in the regions before and after the MS. As already discussed for the numerical results, the large reduction in the amplitude ratio in the frequency band $3.55-4.55$ kHz (see the black dashed box in Fig.~\ref{fig10}d) identifies the region in which the nonlocal metasurface is effective and provides broadband wave blocking effect. Note that, although the operating range is slightly smaller than the one predicted via numerical simulations, it still covers a range of 1 kHz, that is approximately a $24.69\%$ relative bandwidth with respect to the center frequency $f_0$=4.05 kHz. Compared to the numerical results, we observe a reduction in the operating range and a global shift towards lower frequencies. These differences are due to structural modifications applied to the original design during the fabrication phase, to some variability in the properties of the 3D printed connecting beams, to the coarseness of the printing process, and to inaccuracies during the assembly procedure of the nonlocal links (which where glued on location). Despite these differences, the experimental results still show a marked increase in the operating range that is approximately six times that observed in local MS.

\begin{figure}[ht]
    \centering
	\includegraphics[scale=0.7]{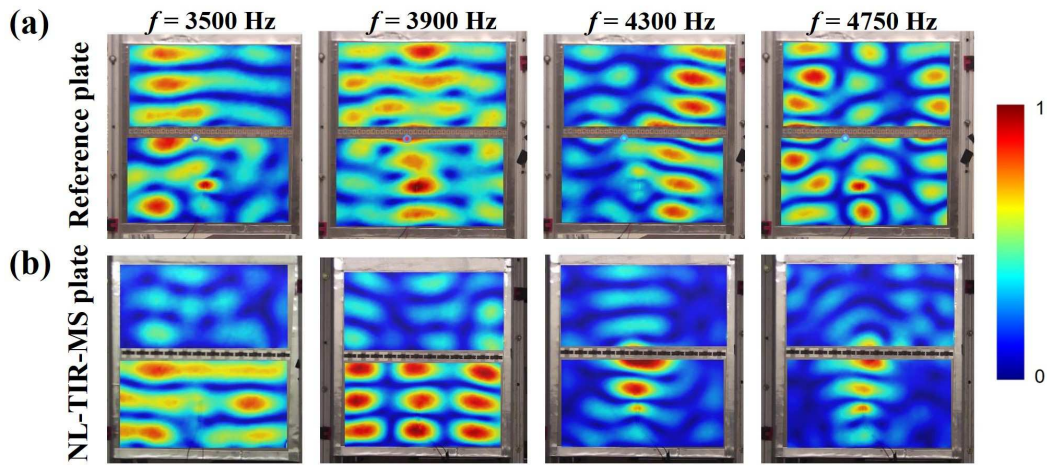}
	\caption{Full field experimental data showing the performance of the nonlocal TIR-MS in comparison to the local design. The images show the out-of-plane particle velocity field at four selected frequencies within the operating bandwith of the nonlocal MS. (a) the out-of-plane particle velocity amplitude for the reference (local) MS configuration. (b) the out-of-plane particle velocity amplitude for the nonlocal TIR-MS. The comparison of the wave patterns clearly indicates the broadband wave blocking effect of the non-local TIR-MS compared to a local design.} \label{fig11}
\end{figure}

Figure \ref{fig11} summarizes the results of the experimental measurements in terms of full field data. In particular, Fig.~\ref{fig11}a and b show the response of both the reference and the NL-TIR-MS waveguides at four selected frequencies ($f_1=3.5$, $f_2=3.9$, $f_3=4.3$, and $f_4=4.75$ kHz, see the four yellow dashed line in Fig.~\ref{fig10}b as well) within the operating range of the nonlocal metasurface. The response was measured in terms of out-of-plane ($z-$direction) velocity $V_{z}$ amplitude field. As expected, in the nonlocal TIR-MS waveguide the vibrational energy is mostly confined in the area preceding the MS, while only a small fraction propagates through the MS interface. On the other hand, in the reference (local) waveguide, the wave could penetrate the MS practically unaffected. The direct comparison of these wave patterns further support the validity of the intentionally nonlocal concept to achieve passive broadband metasurfaces.

\section{Conclusions}
We have presented and experimentally validated the concept of intentional nonlocality in order to design passive elastic metasurfaces capable of broadband performance. 
Contrarily to the classical metasurface design, where wave control is achieved by engineering the phase gradient via local resonances, in the nonlocal design the response at a point of the metasurface depends simultaneously on the response of multiple distant points along the interface. This unique behavior was possible due to the use of specifically designed connecting elements that provided carefully tuned effective dynamic properties and coupled together multiple distant units. In essence, these coupling elements represent a macroscopic implementation of the concept of action-at-a-distance typical of nonlocal microscopic strictures.

Dedicated modeling tools were also developed in order account for the nonlocal coupling forces between unit cells and to characterize the dynamic behavior of nonlocal supercells. As a practical example, the methodology was applied to the design of a broadband total internal reflection metasurface. The performance of this design was validated via both full field numerical simulations and experimental testing. Both results confirmed the validity of the nonlocal concept and the ability to achieve broadband performance via a fully passive approach.

While the nonlocal design was tested on a TIR type metasurface, the concept is very general and it is expected to be applicable to any kind of metasurface. The ability to overcome the intrinsic narrowband nature of traditional metasurface designs is expected to eliminate a significant barrier to the use of these passive wave-control devices and to pave the way to a wide range of practical applications.

\section*{Acknowledgements}
The authors gratefully acknowledge the financial support of the Sandia National Laboratory under the Academic Alliance program, grant \#1847039. The authors thank Mr. Dylan Casey and Mr. David Saiz for their fundamental contribution on the metal printing process of the nonlocal units. H.Z. and F.S. gratefully acknowledge the National Science Foundation for partial financial support under grant \#1761423.

\section*{Supplementary information}
Please see the methodology details, additional numerical results, and the design's geometric details in SI\cite{SI}.

\bibliography{sample}

\end{document}